\documentstyle[amsfonts,12pt,thmsa,sw20lart]{article}

\input tcilatex

\begin{document}

\author{Emil Prodan \\
University of Houston\\
4800 Calhoun Rd\\
Houston Tx 77204-5508\\
e-mail: prodan@shasta.phys.uh.edu}
\title{The Laplace-Beltrami operator on surfaces with axial symmetry}
\maketitle

\begin{abstract}
A solution for the mathematical problem of functional calculus with
Laplace-Beltrami operator on surfaces with axial symmetry is found. A
quantitative analysis of the spectrum is presented.
\end{abstract}

\section{Introduction}

The physical situation which has initiated this research is that of a
dielectric particle with electric charges on its surface, placed in electric
field. Here, the diffusion equation of the charges is coupled with the
Maxwell equations. There is an analytical solution of this system of
equation \cite{PR} which involves some functional calculus with operators,
in particular with Laplace-Beltrami operator defined on the surface of that
particle. We can imagine many other physical situations described by a
complicated system of equations where the Laplace-Beltrami operator is
implicated (e.g. that of the acoustic wave scattering on particles with
membrane, etc.). As before, one can find a compact solution by using
functional calculus. However, these solutions are not complete because, at
this level, all is formal. We must have an effective procedure to calculate
the expressions which involve operators. One can try to compute the matrices
of those operators in some orthonormal basis and to transform the problem in
to algebraic one. The practical problem is that one can compute only a
finite number of matrix elements and this can leads to serious problem when
unbounded operators are implicated. If we choose an inappropriate basis, it
is possible that the expressions, calculated with truncated matrices, to not
converge at the correct result.

In this paper we will find an orthonormal basis in the space of square
integrable functions defined on a surfaces with axial symmetry such that the
truncated matrices of Laplace-Beltrami operator converge in the norm
resolvent sense. Then, according to \cite{RS1} we can use these truncated
matrices in functional calculus.

\section{The result}

Let ${\bf M}$ be a $C^\infty $ closed 2-dimensional surface which in the
spherical coordinates $\left\{ r,\theta ,\phi \right\} $ relative to a
3-orthogonal system of axes is defined by the equation $r=r\left( \theta
\right) $. We consider that all necessary conditions to have a $C^\infty $
surface are fulfilled. Let this surface be equipped with the metric which is
induced by the embedding in ${\bf R}^3$ and let $x_0\in {\bf M}$ be the
point defined by $\theta =0$. Relative to this point, the normal coordinates 
$\left\{ \lambda ,\varphi \right\} $ are defined by 
\[
\left\{ 
\begin{array}{c}
\lambda \left( x\left( \theta ,\phi \right) \right) =d\left( x\left( \theta
,\phi \right) ,x_0\right) =\stackrel{\theta }{\stackunder{0}{\dint }}dt\sqrt{%
r\left( t\right) ^2+r^{\prime }\left( t\right) ^2} \\ 
\varphi =\phi
\end{array}
\right. 
\]
which parameterize the entire surface, without the points $\theta =0,\pi $.
We define $R=\lambda \left( x\left( \pi \right) \right) /\pi $ and the new
coordinates: $\left\{ \vartheta =\lambda /R,\ \varphi \right\} $. In these
coordinates, the metric form is 
\[
g\left( \vartheta ,\varphi \right) =\left( 
\begin{array}{cc}
R^2 & 0 \\ 
0 & r\left( \theta \left( \vartheta \right) \right) ^2\sin \left( \theta
\left( \vartheta \right) \right) ^2
\end{array}
\right) \text{.} 
\]

\smallskip\ 

\begin{proposition}
The set of $C^{\infty }$ functions: 
\[
{\cal Y}_{lm}:{\bf M}\rightarrow {\bf C}\text{, }{\cal Y}_{lm}\left(
\vartheta ,\varphi \right) =\sqrt{\frac{R\sin \vartheta }{r\left( \theta
\left( \vartheta \right) \right) \sin \theta \left( \vartheta \right) }}%
\frac{Y_{lm}\left( \vartheta ,\varphi \right) }{R}\text{, }m\in {\bf Z}%
,l\geq \left| m\right| 
\]
is an orthonormal basis in $L_{2}\left( {\bf M},\mu _{g}\right) $, where $%
Y_{lm}$ represent the spherical harmonics and $\mu _{g}$ is the measure
induced on ${\bf M}$ by the metric $g$.

\smallskip\ 
\end{proposition}

{\it Proof. }The orthonormality: 
\[
\left\langle {\cal Y}_{lm},\ {\cal Y}_{l^{\prime }m^{\prime }}\right\rangle =%
\stackrel{\pi }{\stackunder{0}{\int }}d\vartheta \stackrel{2\pi }{%
\stackunder{0}{\int }}d\varphi \ \sqrt{\det g}\cdot {\cal Y}_{lm}\left(
\vartheta ,\varphi \right) {\cal Y}_{l^{\prime }m^{\prime }}^{*}\left(
\vartheta ,\varphi \right) = 
\]
\[
\stackrel{\pi }{\stackunder{0}{\int }}d\vartheta \stackrel{2\pi }{%
\stackunder{0}{\int }}d\varphi \ Rr\left( \theta \right) \sin \theta \cdot 
\frac{R\sin \vartheta }{r\sin \theta }\frac{{\cal Y}_{lm}\left( \vartheta
,\varphi \right) }R\frac{{\cal Y}_{l^{\prime }m^{\prime }}^{*}\left(
\vartheta ,\varphi \right) }R= 
\]
\[
\stackrel{\pi }{\stackunder{0}{\int }}d\vartheta \stackrel{2\pi }{%
\stackunder{0}{\int }}d\varphi \ \sin \vartheta \cdot {\cal Y}_{lm}\left(
\vartheta ,\varphi \right) {\cal Y}_{l^{\prime }m^{\prime }}^{*}\left(
\vartheta ,\varphi \right) =\delta _{ll^{\prime }}\delta _{mm^{\prime }} 
\]

The completeness: 
\[
\stackrel{\pi }{\stackunder{0}{\int }}d\vartheta ^{\prime }\stackrel{2\pi }{%
\stackunder{0}{\int }}d\varphi ^{\prime }\ \sqrt{\det g}\cdot \stackunder{l,m%
}{\sum }{\cal Y}_{lm}\left( \vartheta ,\varphi \right) {\cal Y}_{l^{\prime
}m^{\prime }}^{*}\left( \vartheta ^{\prime },\varphi ^{\prime }\right)
f\left( \vartheta ^{\prime },\varphi ^{\prime }\right) = 
\]
\[
\stackrel{\pi }{\stackunder{0}{\int }}d\vartheta ^{\prime }\stackrel{2\pi }{%
\stackunder{0}{\int }}d\varphi ^{\prime }\ Rr\left( \theta ^{\prime }\right)
\sin \left( \theta ^{\prime }\right) \stackunder{l,m}{\sum }\sqrt{\frac{%
R\sin \vartheta }{r\left( \theta \right) \sin \theta }}\sqrt{\frac{R\sin
\vartheta ^{\prime }}{r\left( \theta ^{\prime }\right) \sin \theta ^{\prime }%
}}\times 
\]
\[
\frac{Y_{lm}\left( \vartheta ,\varphi \right) }R\frac{Y_{lm}^{*}\left(
\vartheta ^{\prime },\varphi ^{\prime }\right) }Rf\left( \vartheta ^{\prime
},\varphi ^{\prime }\right) =\sqrt{\frac{R\sin \vartheta }{r\left( \theta
\right) \sin \theta }}\times 
\]
\[
\stackrel{\pi }{\stackunder{0}{\int }}d\vartheta ^{\prime }\stackrel{2\pi }{%
\stackunder{0}{\int }}d\varphi ^{\prime }\ \stackunder{l,m}{\sum }%
Y_{lm}\left( \vartheta ,\varphi \right) Y_{lm}^{*}\left( \vartheta ^{\prime
},\varphi ^{\prime }\right) \sqrt{\frac{r\left( \theta ^{\prime }\right)
\sin \theta ^{\prime }}{R\sin \left( \vartheta ^{\prime }\right) }}f\left(
\vartheta ^{\prime },\varphi ^{\prime }\right) =f\left( \vartheta ,\varphi
\right) \text{,} 
\]
because $\sqrt{\frac{r\left( \theta \right) \sin \theta }{R\sin \left(
\vartheta \right) }}f\left( \vartheta ,\varphi \right) $ is in $L_2\left( 
{\bf M},\mu _g\right) $ if $f\in L_2\left( {\bf M},\mu _g\right) $.%
\endproof%
%

\smallskip\ 

For a fixed $m$, let ${\bf S}_m$ be the Hilbert subspace spanned by $\left\{ 
{\cal Y}_{lm}\right\} _{l\geq \left| m\right| }$, which is invarianted by
the Laplace Beltrami operator. In the following, we will consider the
restriction of this operator at a ${\bf S}_m$ subspace, $\Delta ^{\left(
m\right) }=\Delta \mid _{{\bf S}_m}$. Let $P_k^{\left( m\right) }$, $k\geq
\left| m\right| $, be the projection on the subspace spanned by the vectors $%
{\cal Y}_{\left| m\right| m}$, ...,${\cal Y}_{km}$. Our main result is:

\smallskip\ 

\begin{theorem}
The sequence of operators 
\[
\left\{ P_k^{\left( m\right) }\left[ P_k^{\left( m\right) }\circ \Delta
^{\left( m\right) }\circ P_k^{\left( m\right) }-z\right] ^{-1}\right\}
_{k\geq \left| m\right| } 
\]
converges in norm topology at the operator $\left[ \Delta ^{\left( m\right)
}-z\right] ^{-1}$, for any $z\in {\bf C}$ with $\func{Im}z\neq 0$.

\smallskip\ 
\end{theorem}

{\it Proof.} We have successively: 
\[
P_k^{\left( m\right) }\frac 1{P_k^{\left( m\right) }\circ \Delta ^{\left(
m\right) }\circ P_k^{\left( m\right) }-z}-\frac 1{\Delta ^{\left( m\right)
}-z}= 
\]
\[
P_k^{\left( m\right) }\left[ \frac 1{P_k^{\left( m\right) }\Delta ^{\left(
m\right) }P_k^{\left( m\right) }-z}-\frac 1{\Delta ^{\left( m\right) }-z}%
\right] -\left( I-P_k^{\left( m\right) }\right) \frac 1{\Delta ^{\left(
m\right) }-z}= 
\]
\[
\frac 1{P_k^{\left( m\right) }\Delta ^{\left( m\right) }P_k^{\left( m\right)
}-z}P_k^{\left( m\right) }\Delta ^{\left( m\right) }\left( I-P_k^{\left(
m\right) }\right) \frac 1{\Delta ^{\left( m\right) }-z}-\left( I-P_k^{\left(
m\right) }\right) \frac 1{\Delta ^{\left( m\right) }-z}= 
\]
\[
\frac z{P_k^{\left( m\right) }\Delta ^{\left( m\right) }P_k^{\left( m\right)
}-z}\left[ I+\frac 1zP_k^{\left( m\right) }\Delta ^{\left( m\right) }\left(
I-P_k^{\left( m\right) }\right) \right] \left( I-P_k^{\left( m\right)
}\right) \frac 1{\Delta ^{\left( m\right) }-z}. 
\]
Without loss of generality we can choose $z=i\omega $, $\omega \in {\bf R}$, 
$\omega \neq 0$. Thus: 
\[
\left\| P_k^{\left( m\right) }\frac 1{P_k^{\left( m\right) }\circ \Delta
^{\left( m\right) }\circ P_k^{\left( m\right) }-z}-\frac 1{\Delta ^{\left(
m\right) }-z}\right\| \leq 
\]
\[
\left( 1+\frac 1\omega \left\| P_k^{\left( m\right) }\Delta ^{\left(
m\right) }\left( I-P_k^{\left( m\right) }\right) \right\| \right) \cdot
\left\| \left( I-P_k^{\left( m\right) }\right) \frac 1{\Delta ^{\left(
m\right) }-i\omega }\right\| 
\]

\smallskip\ 

\begin{lemma}
For $l$, $k\in {\bf N}$, $l$, $k\geq \left| m\right| $, with $l\neq k$: 
\[
\left\langle {\cal Y}_{lm},\Delta ^{\left( m\right) }{\cal Y}%
_{km}\right\rangle =\left\langle {\cal Y}_{lm},h\cdot {\cal Y}%
_{km}\right\rangle \text{,} 
\]
where $h$ is at least a $C^0$ function on $M$. The operators $P_k^{\left(
m\right) }\circ \Delta ^{\left( m\right) }\circ \left( I-P_k^{\left(
m\right) }\right) $ are bounded and their norms satisfy: 
\[
\left\| P_k^{\left( m\right) }\circ \Delta ^{\left( m\right) }\circ \left(
I-P_k^{\left( m\right) }\right) \right\| \leq \left\| h\right\| _\infty 
\text{.} 
\]

\smallskip\ 
\end{lemma}

It follows that 
\[
\left\| P_k^{\left( m\right) }\frac 1{P_k^{\left( m\right) }\circ \Delta
^{\left( m\right) }\circ P_k^{\left( m\right) }-z}-\frac 1{\Delta ^{\left(
m\right) }-z}\right\| \leq 
\]
\[
\left( 1+\frac{\left\| h\right\| _\infty }\omega \right) \cdot \left\|
\left( I-P_k^{\left( m\right) }\right) \frac 1{\Delta ^{\left( m\right)
}-i\omega }\right\| \text{.} 
\]
To evaluate the last norm, we use the following

\smallskip\ 

\begin{lemma}
Let $s\left( \vartheta \right) $ be the quantity $\sqrt{\frac{r\sin \theta }{%
R\sin \vartheta }}$. If $\lambda _{\left| m\right| }^{\left( m\right) }$%
,..., $\lambda _n^{\left( m\right) }$,... are the ordered eigenvalues of $%
\Delta ^{\left( m\right) }$ and $v_{\left| m\right| }^{\left( m\right) }$%
,..., $v_n^{\left( m\right) }$,...are the corresponding eigenvectors, then
for any $m\in {\bf Z}$ and $l\geq \left| m\right| $, $l\geq 1$ and $n\geq
\left| m\right| $%
\[
\left| \left\langle {\cal Y}_{lm}\mid v_n^{\left( m\right) }\right\rangle
\right| \leq \frac{Rc\left[ \left\| ds\right\| _\infty +c\sqrt{\lambda
_n^{\left( m\right) }}\right] }{\sqrt{l\left( l+1\right) }}\text{,} 
\]
and for any $m\in {\bf Z}$, $l\geq \left| m\right| $, and $n\geq \left|
m\right| $, $n\geq 1$ 
\[
\left| \left\langle {\cal Y}_{lm}\mid v_n^{\left( m\right) }\right\rangle
\right| \leq \frac{c\left[ \left\| ds^{-1}\right\| _\infty +\frac cR\sqrt{%
l\left( l+1\right) }\right] }{\sqrt{\lambda _n^{\left( m\right) }}}\text{.} 
\]
Moreover: 
\[
\frac 1{c^2}\frac{l\left( l+1\right) }{R^2}\leq \lambda _l^{\left( m\right)
}\leq c^2\frac{l\left( l+1\right) }{R^2}\text{,} 
\]
where $c=\left[ \max \left\{ \left\| s^2\right\| _\infty ,\left\|
s^{-2}\right\| _\infty \right\} \right] ^{1/2}$.

\smallskip\ 
\end{lemma}

Now, let $v\in L_2\left( {\bf M},\mu _g\right) $, $v=\stackunder{n\geq
\left| m\right| }{\sum }a_n\cdot v_n^{\left( m\right) }$. Then 
\[
\left\| \left( I-P_k^{\left( m\right) }\right) \frac 1{\Delta ^{\left(
m\right) }-i\omega }v\right\| ^2=\left\| \stackunder{l\geq k+1}{\sum }\ 
\stackunder{n\geq \left| m\right| }{\sum }a_n\cdot \frac{\left\langle {\cal Y%
}_{lm},v_n^{\left( m\right) }\right\rangle }{\lambda _n^{\left( m\right)
}-i\omega }{\cal Y}_{lm}\right\| ^2= 
\]
\[
\stackunder{l\geq k+1}{\sum }\left| \stackunder{n\geq \left| m\right| }{\sum 
}a_n\cdot \frac{\left\langle {\cal Y}_{lm},v_n^{\left( m\right)
}\right\rangle }{\lambda _n^{\left( m\right) }-i\omega }\right| ^2\leq 
\]
\[
\stackunder{l\geq k+1}{\sum }\ \stackunder{n\geq \left| m\right| }{\sum }%
\left| a_n\right| ^2\cdot \stackunder{n\geq \left| m\right| }{\sum }\left| 
\frac{\left\langle {\cal Y}_{lm},v_n^{\left( m\right) }\right\rangle }{%
\lambda _n^{\left( m\right) }-i\omega }\right| ^2\leq 
\]
$\ $%
\[
\left\| v\right\| ^2\stackunder{l\geq k+1}{\sum }\ \stackunder{n\geq \left|
m\right| }{\sum }\left| \frac{cR\left[ \left\| ds\right\| _\infty +c\sqrt{%
\lambda _n^{\left( m\right) }}\right] }{\sqrt{l\left( l+1\right) }\left(
\lambda _n^{\left( m\right) }-i\omega \right) }\right| ^2= 
\]
\[
\left( cR\right) ^2\frac{\left\| v\right\| ^2}{k+1}\stackunder{n\geq \left|
m\right| }{\sum }\left| \frac{\left[ \left\| ds\right\| _\infty +c\sqrt{%
\lambda _n^{\left( m\right) }}\right] }{\lambda _n^{\left( m\right)
}-i\omega }\right| ^2. 
\]

For $\left| m\right| >0$, 
\[
\left\| \left( I-P_k^{\left( m\right) }\right) \frac 1{\Delta ^{\left(
m\right) }-i\omega }v\right\| ^2\leq 
\]
\[
\left( cR\right) ^2\frac{\left\| v\right\| ^2}{k+1}\stackunder{n\geq \left|
m\right| }{\sum }\frac 1{\lambda _n^{\left( m\right) }}\frac{\left[ c+\frac{%
\left\| ds\right\| _\infty }{\sqrt{\lambda _n^{\left( m\right) }}}\right] ^2%
}{1+\frac{\omega ^2}{\lambda _n^{\left( m\right) 2}}}\leq 
\]
\[
\left( cR\right) ^2\frac{\left\| v\right\| ^2}{k+1}\left[ c+\frac{\left\|
ds\right\| _\infty }{\sqrt{\lambda _{\left| m\right| }^{\left( m\right) }}}%
\right] ^2\stackunder{n\geq \left| m\right| }{\sum }\frac 1{\lambda
_n^{\left( m\right) }}. 
\]
Finally 
\[
\left\| \left( I-P_k^{\left( m\right) }\right) \frac 1{\Delta ^{\left(
m\right) }-i\omega }\right\| \leq \frac{\left( cR\right) ^2}{\sqrt{\left(
k+1\right) \left| m\right| }}\left[ c+\frac{cR\left\| ds\right\| _\infty }{%
\sqrt{\left| m\right| \left( \left| m\right| +1\right) }}\right] \text{.} 
\]

For $\left| m\right| =0$, 
\[
\left\| \left( I-P_k^{\left( 0\right) }\right) \frac 1{\Delta ^{\left(
0\right) }-i\omega }v\right\| ^2\leq 
\]
\[
\frac{\left\| v\right\| ^2}{k+1}\left( \frac{\left( cR\right) ^2\left\|
ds\right\| _\infty ^2}{\omega ^2}+\stackunder{n\geq 1}{\sum }\left(
cR\right) ^2\frac 1{\lambda _n^{\left( 0\right) }}\frac{\left[ c+\frac{%
\left\| ds\right\| _\infty }{\sqrt{\lambda _n^{\left( 0\right) }}}\right] ^2%
}{1+\frac{\omega ^2}{\lambda _n^{\left( 0\right) 2}}}\right) \leq 
\]
\[
\left( cR\right) ^2\frac{\left\| v\right\| ^2}{k+1}\left( \frac{\left\|
ds\right\| _\infty ^2}{\omega ^2}+\left( cR\right) ^2\left[ c+\frac{%
cR\left\| ds\right\| _\infty }{\sqrt{2}}\right] ^2\right) 
\]
thus: 
\[
\left\| \left( I-P_k^{\left( 0\right) }\right) \frac 1{\Delta ^{\left(
0\right) }-i\omega }\right\| \leq \frac{\left( cR\right) ^2}{\sqrt{\left(
k+1\right) }}\sqrt{\frac{\left\| ds\right\| _\infty ^2}{\left( cR\right)
^2\omega ^2}+\left[ c+\frac{cR\left\| ds\right\| _\infty }{\sqrt{2}}\right]
^2}\text{.} 
\]
Having that $\left\| ds\right\| _\infty =\frac 1R\left\| \frac{\partial s}{%
\partial \vartheta }\right\| _\infty =\frac 1R\left\| s^{\prime }\right\|
_\infty $ we can conclude: 
\[
\left\| P_k^{\left( m\right) }\frac 1{P_k^{\left( m\right) }\circ \Delta
^{\left( m\right) }\circ P_k^{\left( m\right) }-i\omega }-\frac 1{\Delta
^{\left( m\right) }-i\omega }\right\| \leq 
\]
\[
\left\{ 
\begin{array}{c}
\left( 1+\frac{\left\| h\right\| _\infty }\omega \right) \frac{\left(
cR\right) ^2}{\sqrt{\left( k+1\right) }}\sqrt{\frac{\frac 1R\left\|
s^{\prime }\right\| _\infty }{\left( cR\right) ^2\omega ^2}+c^2\left[ 1+%
\frac{\left\| s^{\prime }\right\| _\infty }{\sqrt{2}}\right] ^2}\text{, for }%
m=0 \\ 
\\ 
\left( 1+\frac{\left\| h\right\| _\infty }\omega \right) \frac{c^3R^2}{\sqrt{%
\left( k+1\right) \left| m\right| }}\left[ 1+\frac{\left\| s^{\prime
}\right\| _\infty }{\sqrt{\left| m\right| \left( \left| m\right| +1\right) }}%
\right] \text{, for }\left| m\right| \geq 1
\end{array}
\right. 
\]

\endproof%
%

\smallskip\ 

{\it Proof of Lemma 3. }We have successively 
\[
\left\langle {\cal Y}_{lm},\Delta ^{\left( m\right) }{\cal Y}%
_{km}\right\rangle =\left\langle d{\cal Y}_{lm},d{\cal Y}_{km}\right\rangle
= 
\]
\[
\stackrel{\pi }{\stackunder{0}{\int }}d\vartheta \stackrel{2\pi }{%
\stackunder{0}{\int }}d\varphi \ R^2\sin \vartheta s^2\left[ \frac
1{R^2}\frac \partial {\partial \vartheta }\left( \frac{Y_{lm}^{\star }}{sR}%
\right) \frac \partial {\partial \vartheta }\left( \frac{Y_{km}}{sR}\right) +%
\frac{m^2s^{-6}}{R^2\sin \vartheta ^2}\frac{Y_{lm}^{\star }}R\frac{Y_{km}}R%
\right] = 
\]
$\ $%
\[
\stackrel{\pi }{\stackunder{0}{\int }}d\vartheta \stackrel{2\pi }{%
\stackunder{0}{\int }}d\varphi \frac{\sin \vartheta }{R^2}\left[ \frac
\partial {\partial \vartheta }Y_{lm}^{\star }\frac \partial {\partial
\vartheta }Y_{km}-\frac{\partial \ln s}{\partial \vartheta }\frac \partial
{\partial \vartheta }\left( Y_{lm}^{\star }Y_{km}\right) +\left( \frac{%
\partial \ln s}{\partial \vartheta }\right) ^2Y_{lm}^{\star }Y_{km}\right] + 
\]
\[
\frac 1{R^2}\stackrel{\pi }{\stackunder{0}{\int }}d\vartheta \stackrel{2\pi 
}{\stackunder{0}{\int }}d\varphi \sin \left( \vartheta \right) \frac{%
m^2s^{-4}}{R^2\sin ^2\vartheta }Y_{lm}^{\star }Y_{km}= 
\]
$\ $%
\[
\frac 1{R^2}\stackrel{\pi }{\stackunder{0}{\int }}d\vartheta \stackrel{2\pi 
}{\stackunder{0}{\int }}d\varphi \sin \vartheta \left[ \frac \partial
{\partial \vartheta }Y_{lm}^{\star }\frac \partial {\partial \vartheta
}Y_{km}+\frac{m^2s^{-4}}{R^2\sin ^2\vartheta }Y_{lm}^{\star }Y_{km}\right] + 
\]
\[
\frac 1{R^2}\stackrel{\pi }{\stackunder{0}{\int }}d\vartheta \stackrel{2\pi 
}{\stackunder{0}{\int }}d\phi \sin \vartheta \left[ \left( \frac{\partial
\ln s}{\partial \vartheta }\right) ^2-\frac 1{\sin \vartheta }\frac \partial
{\partial \vartheta }\left( \sin \vartheta \frac{\partial \ln s}{\partial
\vartheta }\right) \right] \cdot Y_{lm}^{\star }Y_{km}= 
\]
$\ $%
\[
\stackrel{\pi }{\stackunder{0}{\int }}d\vartheta \stackrel{2\pi }{%
\stackunder{0}{\int }}d\varphi \frac{\sin \vartheta }{R^2}\left[ m^2\frac{%
s^{-4}-1}{R^2\sin ^2\vartheta }+\left( \frac{\partial \ln s}{\partial
\vartheta }\right) ^2-\frac 1{\sin \vartheta }\frac \partial {\partial
\vartheta }\left( \sin \vartheta \frac{\partial \ln s}{\partial \vartheta }%
\right) \right] Y_{lm}^{\star }Y_{km}\text{.} 
\]
It is easy to check that $s\left( \vartheta \right) $ is at least of $C^2$
class, so the function $h:\left[ 0,\pi \right] \rightarrow {\bf R}$, 
\[
h\left( \vartheta \right) =m^2\frac{s^{-4}-1}{R^2\sin ^2\vartheta }+\left( 
\frac{\partial \ln s}{\partial \vartheta }\right) ^2-\frac 1{\sin \vartheta
}\frac \partial {\partial \vartheta }\left( \sin \vartheta \frac{\partial
\ln s}{\partial \vartheta }\right) 
\]
is at least of $C^0$ class. Finally 
\[
\left\langle {\cal Y}_{lm},\Delta ^{\left( m\right) }{\cal Y}%
_{km}\right\rangle =\frac 1{R^2}\stackrel{\pi }{\stackunder{0}{\int }}%
d\vartheta \stackrel{2\pi }{\stackunder{0}{\int }}d\phi \sin \vartheta \cdot
h\left( \vartheta \right) \cdot Y_{lm}^{\star }Y_{km}= 
\]
\[
\stackrel{\pi }{\stackunder{0}{\int }}d\vartheta \stackrel{2\pi }{%
\stackunder{0}{\int }}d\varphi \sin \left( \vartheta \right) s^2\cdot
h\left( \vartheta \right) \cdot \frac{Y_{lm}^{\star }}{Rs}\frac{Y_{km}}{Rs}%
=\left\langle {\cal Y}_{lm},h\cdot {\cal Y}_{km}\right\rangle . 
\]

For the second part 
\[
\left| \left\langle v\left| P_k^{\left( m\right) }\circ \Delta ^{\left(
m\right) }\circ \left( I-P_k^{\left( m\right) }\right) \right|
u\right\rangle \right| =\left| \left\langle P_k^{\left( m\right) }v\left|
\Delta ^{\left( m\right) }\right| \left( I-P_k^{\left( m\right) }\right)
u\right\rangle \right| = 
\]
\[
\left| \left\langle P_k^{\left( m\right) }v,h\cdot \left( I-P_k^{\left(
m\right) }\right) u\right\rangle \right| \leq \left\| P_k^{\left( m\right)
}v\right\| \cdot \left\| h\cdot \left( I-P_k^{\left( m\right) }\right)
u\right\| \leq \left\| h\right\| _\infty \left\| v\right\| \cdot \left\|
u\right\| . 
\]

\endproof%
%

\smallskip\ 

{\it Proof of Lemma 4.}

\smallskip\ 

\begin{proposition}
The application $\tilde{g}:\left( 0,\pi \right) \times \left[ 0,2\pi \right]
\longrightarrow M\left( 2\times 2\right) $%
\[
\tilde{g}\left( \vartheta ,\varphi \right) =\left( 
\begin{array}{cc}
1 & 0 \\ 
0 & R^2\sin \left( \vartheta \right) ^2
\end{array}
\right) , 
\]
\end{proposition}

defines a metric on ${\bf M}$. Moreover, $\sqrt{\frac{\det g}{\det \tilde{g}}%
}=s^2$.

\smallskip\ 

If we consider the spaces of the squared integrable functions with the
measures induced by the two metrics, $L_2\left( {\bf M},\mu _g\right) $ and $%
L_2\left( {\bf M},\mu _{\tilde{g}}\right) $, and the spaces of
one-differential forms with the standard scalar products, $A^{\left(
1\right) }\left( {\bf M},\mu _g\right) $ and $A^{\left( 1\right) }\left( 
{\bf M},\mu _{\tilde{g}}\right) $, then:

\smallskip\ 

\begin{proposition}
The spaces $L_2\left( {\bf M},\mu _g\right) $and $L_2\left( {\bf M},\mu _{%
\tilde{g}}\right) $ coincide, $A^{\left( 1\right) }\left( {\bf M},\mu
_g\right) $ and $A^{\left( 1\right) }\left( {\bf M},\mu _{\tilde{g}}\right) $
coincide too.

\smallskip\ 
\end{proposition}

{\it Proof.} For $f\in L_2\left( {\bf M},\mu _g\right) $ we have: 
\[
\left\| f\right\| _{\tilde{g}}=\stackrel{\pi }{\stackunder{0}{\int }}%
d\vartheta \stackrel{2\pi }{\stackunder{0}{\int }}d\varphi \sqrt{\det \tilde{%
g}\left( \vartheta \right) }\left| f\left( \vartheta ,\varphi \right)
\right| ^2\leq \left\| \frac{\sqrt{\det \tilde{g}\left( \vartheta \right) }}{%
\sqrt{\det g\left( \vartheta \right) }}\right\| _\infty \cdot \left\|
f\right\| _g^2\leq \infty \text{,} 
\]
thus: $f\in L_2\left( {\bf M},\mu _{\tilde{g}}\right) .$Analogous, for $f\in
L_2\left( {\bf M},\mu _{\tilde{g}}\right) $ results: 
\[
\left\| f\right\| _g=\stackrel{\pi }{\stackunder{0}{\int }}d\vartheta 
\stackrel{2\pi }{\stackunder{0}{\int }}d\varphi \sqrt{\det g\left( \vartheta
\right) }\left| f\left( \vartheta ,\varphi \right) \right| ^2\leq \left\| 
\frac{\sqrt{\det g\left( \vartheta \right) }}{\sqrt{\det \tilde{g}\left(
\vartheta \right) }}\right\| _\infty \cdot \left\| f\right\| _{\tilde{g}%
}^2\leq \infty \text{,} 
\]
thus $f\in L_2\left( {\bf M},\mu _g\right) $.

\smallskip\ 

Let $\omega \in A^{\left( 1\right) }\left( {\bf M},\mu _g\right) $, $\omega
=\omega _\vartheta d\vartheta +\omega _\varphi d\varphi $. Will follow 
\[
\stackrel{\pi }{\stackunder{0}{\int }}d\vartheta \stackrel{2\pi }{%
\stackunder{0}{\int }}d\varphi \sqrt{\det \tilde{g}\left( \vartheta \right) }%
g\left( \bar{\omega},\omega \right) =\stackrel{\pi }{\stackunder{0}{\int }}%
d\vartheta \stackrel{2\pi }{\stackunder{0}{\int }}d\varphi \sqrt{\det \tilde{%
g}\left( \vartheta \right) }\left[ \left| \omega _\vartheta \right| ^2+\frac{%
\left| \omega _\varphi \right| ^2}{\det \tilde{g}}\right] \leq 
\]
\[
\leq \max \left\{ \sqrt{\frac{\det \tilde{g}}{\det g}},\sqrt{\frac{\det g}{%
\det \tilde{g}}}\right\} \cdot \left\| \omega \right\| _g^2\leq \infty \text{%
,} 
\]
thus $\omega \in A^{\left( 1\right) }\left( {\bf M},\mu _{\tilde{g}}\right) $%
. The same steps can be followed to show that $\omega \in A^{\left( 1\right)
}\left( {\bf M},\mu _{\tilde{g}}\right) \Rightarrow \omega \in A^{\left(
1\right) }\left( {\Bbb M},\mu _g\right) $. Denoting 
\[
c=\sqrt{\max \left\{ \left\| s^2\right\| _\infty ,\left\| s^{-2}\right\|
_\infty \right\} }\text{,} 
\]
we have on $L_2\left( {\bf M},\mu _g\right) \equiv L_2\left( {\bf M},\mu _{%
\tilde{g}}\right) $: 
\[
\frac 1c\left\| \ \right\| _{\tilde{g}}\leq \left\| \ \right\| _g\leq
c\left\| \ \right\| _{\tilde{g}}\text{,} 
\]
and, on $A^{\left( 1\right) }\left( {\bf M},\mu _{\tilde{g}}\right) \equiv
A^{\left( 1\right) }\left( {\bf M},\mu _g\right) $: 
\[
\frac 1c\left\| \ \right\| _{\tilde{g}}\leq \left\| \ \right\| _g\leq
c\left\| \ \right\| _{\tilde{g}}\text{.} 
\]
\endproof%
%

\smallskip\ 

Now, we have successively 
\[
\left| \left\langle {\cal Y}_{lm}\mid v_n^{\left( m\right) }\right\rangle
_g\right| =\left| \left\langle \frac{Y_{lm}}{sR}\mid v_n^{\left( m\right)
}\right\rangle _g\right| =\left| \left\langle \frac{Y_{lm}}R\mid s^{-1}\cdot
v_n^{\left( m\right) }\right\rangle _g\right| = 
\]
\[
\frac{\left| \left\langle \tilde{\Delta}\frac{Y_{lm}}R\mid s^{-1}\cdot
v_n^{\left( m\right) }\right\rangle _g\right| }{\frac{l\left( l+1\right) }{%
R^2}}=\frac{\left| \left\langle \tilde{\Delta}\frac{Y_{lm}}R\mid s\cdot
v_n^{\left( m\right) }\right\rangle _{\tilde{g}}\right| }{\frac{l\left(
l+1\right) }{R^2}}\leq \frac{\left\| d\frac{Y_{lm}}R\right\| _{\tilde{g}%
}\cdot \left\| d\left( s\cdot v_n^{\left( m\right) }\right) \right\| _{%
\tilde{g}}}{\frac{l\left( l+1\right) }{R^2}}\leq 
\]
$\ $%
\[
\frac R{\sqrt{l\left( l+1\right) }}\left[ \left\| ds\right\| _\infty \left\|
v_n^{\left( m\right) }\right\| _{\tilde{g}}+\left\| s\right\| _\infty
\left\| dv_n^{\left( m\right) }\right\| _{\tilde{g}}\right] \leq \frac{cR%
\left[ \left\| ds\right\| _\infty +c\sqrt{\lambda _n^{\left( m\right) }}%
\right] }{\sqrt{l\left( l+1\right) }}. 
\]

For the second set of inequalities: 
\[
\left| \left\langle {\cal Y}_{lm}\mid v_{n}^{\left( m\right) }\right\rangle
\right| =\frac{1}{\lambda _{n}^{\left( m\right) }}\left| \left\langle {\cal Y%
}_{lm}\mid \Delta v_{n}^{\left( m\right) }\right\rangle \right| \leq \frac{1%
}{\lambda _{n}^{\left( m\right) }}\left\| d{\cal Y}_{lm}\right\| _{g}\left\|
dv_{n}^{\left( m\right) }\right\| _{g}\leq 
\]
\[
\frac{c\left\| d{\cal Y}_{lm}\right\| _{\tilde{g}}}{\sqrt{\lambda
_{n}^{\left( m\right) }}}\leq \frac{c\left[ \left\| ds^{-1}\right\| _{\infty
}+\frac{c}{R}\sqrt{l\left( l+1\right) }\right] }{\sqrt{\lambda _{n}^{\left(
m\right) }}}. 
\]

For the last set of inequalities of lemma 4, once we have the results of the
last proposition we can follow the way of \cite{BA}, or that presented in 
\cite{CR}.%
\endproof%
%

\section{Numerical application}

In general, the spectrum of the truncated matrices does not converge at the
exact spectrum. Without additional results, one knows that only the lowest
eigenvalue of the truncated matrices converges at the exact value. About
these facts, one can consult \cite{RS2}. The results of the last section
have another important consequence: in the proposed basis, the spectrum of
the truncated matrices converges at the exact spectrum. Moreover, because
the matrix of the Laplace-Beltrami operator in the ${\cal Y}$ basis is
''cvasidiagonal'' in the sense that all nondiagonal elements are bounded by $%
\left\| h\right\| _{\infty }$ and the diagonal elements increase
approximatively as $l\left( l+1\right) /R^{2}$, it is to be expected that
the spectrum of these truncated matrices to be very stable. That means, that
even for low dimensions these matrices give us a good approximation of the
exact spectrum. Let us choose the following particular surfaces for our
numerical application: $r\left( \theta \right) =1+1.2\cos \left( \theta
\right) +3\cos \left( \theta \right) ^{2}$, presented in figure 1. The
eigenvalues for different truncated matrices and $m=0$ are presented in
figures 2-6.

Now, let us choose an orthonormal basis for which the affirmation of Lemma 4
is not true. If $d\mu _g\left( \theta ,\phi \right) =\sigma \left( \theta
,\phi \right) \sin \theta \cdot d\theta d\phi $ is the measure induced by
the metric $g$ in the coordinates $\left\{ \theta ,\phi \right\} $, then:

\begin{proposition}
The set $\stackrel{\sim }{{\cal Y}}$ of functions: 
\[
\stackrel{\sim }{{\cal Y}}_{lm}\left( \theta ,\phi \right) =\frac{%
Y_{lm}\left( \theta ,\phi \right) }{\sqrt{\sigma \left( \theta ,\phi \right) 
}},m=0,1,...,l=\left| m\right| ,\left| m\right| +1,..., 
\]

is an orthonormal basis in $L_2\left( {\bf M},\mu _g\right) $.

\smallskip\ 
\end{proposition}

\noindent The proof of this proposition is analogous with that of
Proposition 1. The eigenvalues of different truncated matrices, calculated
in this basis and for the case $m=0$, are presented in figures 7-11. The
numerical application shows that in this case the spectrum of the truncated
matrices are very unstable. This instability can be considered as an
indicator of the fact that for the $\stackrel{\sim }{{\cal Y}}$ basis the
affirmation of our theorem is not true.

\section{Conclusion}

This paper has shown how to construct an orthonormal basis in the space of
square integrable functions defined on a $C^\infty $ surfaces with axial
symmetry, basis which is appropriate for the problems which involve the
Laplace-Beltrami operator. The procedure is standard, in the sense that it
can be applied following the same steps for any $C^\infty $ surface with
axial symmetry. The stability of the truncated matrices spectrum was
theoretically anticipated and numerically verified. By practical point of
view, this allow us to use truncated matrices with small number of rows and
columns.

\section{List of figures}

\noindent {\bf Fig.1} The particular surface chosen for our numerical
application: $r\left( \theta \right) =1+1.2\cos \left( \theta \right) +3\cos
\left( \theta \right) ^{2}$.

\noindent {\bf Fig.2} The eigenvalues of the 15$\times $15 truncated matrix
of Laplace-Beltrami operator in ${\cal Y}$ orthonormal basis.

\noindent {\bf Fig.3} The eigenvalues of the 20$\times $20 truncated matrix
of Laplace-Beltrami operator in ${\cal Y}$ orthonormal basis.

\noindent {\bf Fig.4} The eigenvalues of the 25$\times $25 truncated matrix
of Laplace-Beltrami operator in ${\cal Y}$ orthonormal basis.

\noindent {\bf Fig.5} The eigenvalues of the 30$\times $30 truncated matrix
of Laplace-Beltrami operator in ${\cal Y}$ orthonormal basis.

\noindent {\bf Fig.6} The superposition of figures 2-5.

\noindent {\bf Fig.7} The eigenvalues of the 15$\times $15 truncated matrix
of Laplace-Beltrami operator in $\stackrel{\sim }{{\cal Y}}$ orthonormal
basis.

\noindent {\bf Fig.8} The eigenvalues of the 20$\times $20 truncated matrix
of Laplace-Beltrami operator in $\stackrel{\sim }{{\cal Y}}$ orthonormal
basis.

\noindent {\bf Fig.9} The eigenvalues of the 25$\times $25 truncated matrix
of Laplace-Beltrami operator in $\stackrel{\sim }{{\cal Y}}$ orthonormal
basis.

\noindent {\bf Fig.10} The eigenvalues of the 30$\times $30 truncated matrix
of Laplace-Beltrami operator in $\stackrel{\sim }{{\cal Y}}$ orthonormal
basis.

\noindent {\bf Fig.11} The superposition of figures 7-10.

\noindent Note: the horizontal coordinate in figures 2-11 is just an
ordering index which puts the eigenvalues of the truncated matrices in an
increasing order.

\FRAME{itbpF}{3.1704in}{2.6455in}{0in}{}{}{figure1.gif}{\special{language
"Scientific Word";type "GRAPHIC";display "USEDEF";valid_file "F";width
3.1704in;height 2.6455in;depth 0in;original-width 3.1254in;original-height
2.6039in;cropleft "0";croptop "1";cropright "1";cropbottom "0";filename
'figure1.gif';file-properties "XNPEU";}}\medskip 

\FRAME{itbpF}{3.1704in}{2.6455in}{0in}{}{}{figure2.gif}{\special{language
"Scientific Word";type "GRAPHIC";display "USEDEF";valid_file "F";width
3.1704in;height 2.6455in;depth 0in;original-width 3.1254in;original-height
2.6039in;cropleft "0";croptop "1";cropright "1";cropbottom "0";filename
'figure2.gif';file-properties "XNPEU";}}\medskip 

\FRAME{itbpF}{3.1704in}{2.6455in}{0in}{}{}{figure3.gif}{\special{language
"Scientific Word";type "GRAPHIC";display "USEDEF";valid_file "F";width
3.1704in;height 2.6455in;depth 0in;original-width 3.1254in;original-height
2.6039in;cropleft "0";croptop "1";cropright "1";cropbottom "0";filename
'Figure3.gif';file-properties "XNPEU";}}\medskip 

\FRAME{itbpF}{3.1704in}{2.6455in}{0in}{}{}{figure4.gif}{\special{language
"Scientific Word";type "GRAPHIC";display "USEDEF";valid_file "F";width
3.1704in;height 2.6455in;depth 0in;original-width 3.1254in;original-height
2.6039in;cropleft "0";croptop "1";cropright "1";cropbottom "0";filename
'figure4.gif';file-properties "XNPEU";}}\medskip 

\FRAME{itbpF}{3.1704in}{2.6455in}{0in}{}{}{figure5.gif}{\special{language
"Scientific Word";type "GRAPHIC";display "USEDEF";valid_file "F";width
3.1704in;height 2.6455in;depth 0in;original-width 3.1254in;original-height
2.6039in;cropleft "0";croptop "1";cropright "1";cropbottom "0";filename
'figure5.gif';file-properties "XNPEU";}}\medskip 

\FRAME{itbpF}{3.1704in}{2.6455in}{0in}{}{}{figure6.gif}{\special{language
"Scientific Word";type "GRAPHIC";display "USEDEF";valid_file "F";width
3.1704in;height 2.6455in;depth 0in;original-width 3.1254in;original-height
2.6039in;cropleft "0";croptop "1";cropright "1";cropbottom "0";filename
'figure6.gif';file-properties "XNPEU";}}\medskip 

\FRAME{itbpF}{3.1704in}{2.6455in}{0in}{}{}{figure7.gif}{\special{language
"Scientific Word";type "GRAPHIC";display "USEDEF";valid_file "F";width
3.1704in;height 2.6455in;depth 0in;original-width 3.1254in;original-height
2.6039in;cropleft "0";croptop "1";cropright "1";cropbottom "0";filename
'figure7.gif';file-properties "XNPEU";}}\medskip 

\FRAME{itbpF}{3.1704in}{2.6455in}{0in}{}{}{figure8.gif}{\special{language
"Scientific Word";type "GRAPHIC";display "USEDEF";valid_file "F";width
3.1704in;height 2.6455in;depth 0in;original-width 3.1254in;original-height
2.6039in;cropleft "0";croptop "1";cropright "1";cropbottom "0";filename
'figure8.gif';file-properties "XNPEU";}}\medskip 

\FRAME{itbpF}{3.1704in}{2.6455in}{0in}{}{}{figure9.gif}{\special{language
"Scientific Word";type "GRAPHIC";display "USEDEF";valid_file "F";width
3.1704in;height 2.6455in;depth 0in;original-width 3.1254in;original-height
2.6039in;cropleft "0";croptop "1";cropright "1";cropbottom "0";filename
'figure9.gif';file-properties "XNPEU";}}\medskip 

\FRAME{itbpF}{3.1704in}{2.6455in}{0in}{}{}{figure10.gif}{\special{language
"Scientific Word";type "GRAPHIC";display "USEDEF";valid_file "F";width
3.1704in;height 2.6455in;depth 0in;original-width 3.1254in;original-height
2.6039in;cropleft "0";croptop "1";cropright "1";cropbottom "0";filename
'figure10.gif';file-properties "XNPEU";}}\medskip 

\FRAME{itbpF}{3.1704in}{2.6455in}{0in}{}{}{figure11.gif}{\special{language
"Scientific Word";type "GRAPHIC";display "USEDEF";valid_file "F";width
3.1704in;height 2.6455in;depth 0in;original-width 3.1254in;original-height
2.6039in;cropleft "0";croptop "1";cropright "1";cropbottom "0";filename
'figure11.gif';file-properties "XNPEU";}}\medskip 

\end{document}